\documentclass[conference, compsoc]{IEEEtran}
\IEEEoverridecommandlockouts
\usepackage[numbers]{natbib}
\usepackage[hyphens]{url}
\usepackage{booktabs}
\usepackage{amsmath,amssymb,amsfonts}
\usepackage{algorithmic}
\usepackage{graphicx}
\usepackage{textcomp}
\usepackage{xcolor}
\usepackage{cprotect}
\usepackage{listings}
\def\BibTeX{
    {\rm B\kern-.05em{\sc i\kern-.025em b}\kern-.08em
    T\kern-.1667em\lower.7ex\hbox{E}\kern-.125emX}
}

\begin{document}

\title{Malware Classification Using Static Disassembly and Machine Learning}

\author{
    \IEEEauthorblockN{Zhenshuo Chen}
    \IEEEauthorblockA{
        \textit{School of Computing} \\
        \textit{Dublin City University} \\
        Dublin, Ireland \\
        chenzs108@outlook.com
    }
    \and
    \IEEEauthorblockN{Eoin Brophy}
    \IEEEauthorblockA{
        \textit{Insight Centre for Data Analytics} \\
        \textit{Dublin City University} \\
        Dublin, Ireland \\
        eoin.brophy7@mail.dcu.ie
    }
    \and
    \IEEEauthorblockN{Tomas Ward}
    \IEEEauthorblockA{
        \textit{Insight Centre for Data Analytics} \\
        \textit{Dublin City University} \\
        Dublin, Ireland \\
        tomas.ward@dcu.ie
    }
}

\maketitle

\begin{abstract}
Network and system security are incredibly critical issues now.
Due to the rapid proliferation of malware, traditional analysis methods struggle with enormous samples.
In this paper, we propose four easy-to-extract and small-scale features,
including sizes and permissions of Windows PE sections, content complexity, and import libraries, to classify malware families,
and use automatic machine learning to search for the best model and hyper-parameters for each feature and their combinations.
Compared with detailed behavior-related features like API sequences, proposed features provide macroscopic information about malware.
The analysis is based on static disassembly scripts and hexadecimal machine code.
Unlike dynamic behavior analysis, static analysis is resource-efficient and offers complete code coverage, but is vulnerable to code obfuscation and encryption.
The results demonstrate that features which work well in dynamic analysis are not necessarily effective when applied to static analysis.
For instance, API 4-grams only achieve 57.96\% accuracy and involve a relatively high dimensional feature set (5000 dimensions).
In contrast, the novel proposed features together with a classical machine learning algorithm (Random Forest) presents very good accuracy at 99.40\% and the feature vector is of much smaller dimension (40 dimensions).
We demonstrate the effectiveness of this approach through integration in IDA Pro, which also facilitates the collection of new training samples and subsequent model retraining.
\end{abstract}


\section{INTRODUCTION}
Network and system security are incredibly critical issues at the moment.
According to \cite{symantec-report}, 142 million threats were being blocked every day in 2019.
Furthermore, new types of malware are appearing all the time and are increasingly aggressive. For instance, the use of malicious PowerShell scripts increased by 1000\% in the same year.
To make matters worse, anti-anti-virus techniques used by attackers are also steadily improving.
The use of polymorphic engines allows malware developers to mutate existing code while retaining the original functions unchanged. This is achieved, for example, through the use of obfuscation and encryption.
This has now led to a rapid proliferation of malware which traditional analysis methods struggle to cope with as these rely on signature matching and heuristic rules.

A signature is a model or hash that can uniquely identify a file by machine code, essential strings, or sensitive instruction sequences.
A large database stores existing samples' signatures and they will be compared with the signature generated from an unknown file to match.
This technique is easy to implement but sensitive even to tiny code modification.
Furthermore, it is also difficult to recognize totally new malware because its signature has not yet been captured in the database.
Heuristic rules are determined by malware experts after analyzing malicious behaviors.
In general, they need to review code instructions, check memory data changes and record system events to understand each sample.

These traditional methods have the same drawback: unseen samples must be manually analyzed before creating signatures or heuristic rules.
However, analysts cannot review each unknown file in practice.
Machine learning approaches in contrast do not rely on understanding code and malicious behaviors.
After training with a wide range of known samples, such methods can more easily identify potential malware compared to human experts.
Some automatic models have been applied in related fields,
such as malware homology analysis by dynamic fingerprints in \cite{dynamic-fingerprint}, and gray-scale image representation of malware in \cite{malware-image}, which did not require disassembly or code execution.

We adopt a machine learning approach based on static analysis in this work. The primary exploration and experiments of this paper are as follows:

\begin{itemize}
    \item
    Four small-scale features are proposed: import libraries, section sizes, section permissions, and content complexity.
    The feature descriptions are in Section.~\ref{sct:import-library}, Section.~\ref{sct:pe-section-size}, Section.~\ref{sct:pe-section-permission}, and Section.~\ref{sct:content-complexity} respectively.
    Unlike traditional large-scale features like API $n$-grams and opcode $n$-grams, which focus on detailed APIs and assembly instructions,
    these new features are easy-to-extract and provide macroscopic information about malware.
    In the experiments using Random Forest, their combination achieved a maximum of $99.40\%$ accuracy with only 40 dimensions.

    \item
    The API $n$-gram, an efficient dynamic behavior feature usually generated by system event logging, is applied to static analysis.
    The result demonstrates that actual API sequences are hard to extract from disassembly scripts.
    Inaccurate $n$-grams have a substantial negative impact on classification.
    In the specific studies done here, the highest accuracy based on such features is only $57.96\%$ with Random Forest and a 5000-dimensional feature vector.

    \item
    A method of using the classifier in practice is proposed.
    With the help of IDA Pro \cite{ida-pro}, the most popular reverse analysis tool, new training data can be generated from the latest known malware.
    It also provides a Python development kit and using this the classifier proposed here is implemented as an IDA Pro plug-in.
    This allows an analyst using IDA Pro to process a malware sample and perform classification immediately within their workflow.
\end{itemize}

\section{BACKGROUND}

\subsection{Machine Code and Assembly Languages}
CPUs can only process machine code, which of course consists of binary numbers.
However, it is incredibly challenging to program directly with machine code so assembly languages are used instead.
They use a mnemonic to represent each low-level machine code or instruction.
There is a strong correspondence between assembly instructions and an architecture's machine code.
Every assembly language is designed for exactly one specific computer architecture, such as ARM and Intel x86.
The conversion from assembly languages to executable machine code is performed by assemblers, available since the 1950s.

\begin{figure}[htbp]
    \centerline{\includegraphics[width=\linewidth]{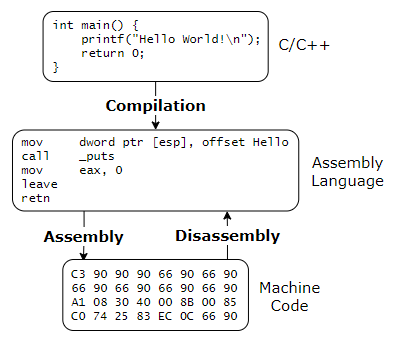}}
    \caption{Compilation, assembly and disassembly}
    \label{fig:assembly}
\end{figure}

Compared with assembly languages, C and C++ are higher-level programming languages.
The source code of C/C++ is translated into assembly languages firstly by compilers, then converted into machine code by assemblers as shown in Fig.~\ref{fig:assembly}.

For malware analysts, there is no high-level source code available for review but only executable files, like \emph{.exe} and \emph{.dll} on Windows systems.
Because of the correspondence between assembly languages and machine code, executable files can be translated into assembly instructions as in Fig.~\ref{fig:assembly}.
This process is called disassembly.

\subsection{Code Obfuscation and Encryption}
Since executable files can be analyzed by disassembly, new techniques called anti-reverse-engineering have been invented to obstruct the process.
Not only malware authors, but also software companies use them to protect commercial software from being cracked or pirated.
Code obfuscation and encryption are two commonly used methods.

Code obfuscation uses needlessly roundabout expressions and data to make source or machine code challenging for humans to understand.
A simple way is to insert data bytes into code. Listing.~\ref{lst:obfuscation} provides an example.
Because of the \verb|jmp next| instruction, the byte definition \verb|db 10| will not be executed as if it does not exist.
But disassemblers may treat this byte as code which makes the following instruction incorrect as Listing.~\ref{lst:obfuscation-ls}.
\verb|db 10| and \verb|mov eax, 0| are wrongly translated as \verb|or bh, byte ptr [eax]|.

\begin{lstlisting}[label=lst:obfuscation, caption=Inserting a data byte into code, language={[x86masm]Assembler}]
    jmp     next
    db      10
next:
    mov     eax, 0
\end{lstlisting}

\begin{lstlisting}[label=lst:obfuscation-ls, caption=Incorrect linear sweep disassembly for Listing.~\ref{lst:obfuscation}, language={[x86masm]Assembler}]
    jmp     next
    or      bh, byte ptr [eax]
\end{lstlisting}

\begin{figure}[htbp]
    \centerline{\includegraphics[width=\linewidth]{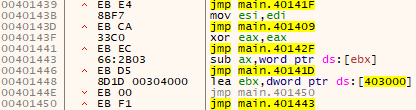}}
    \caption{Execution flow is disrupted by jump instructions}
    \label{fig:obfuscation}
\end{figure}

Another way of performing code obfuscation is to use roundabout expressions.
For example, a logical operation \verb|a^c| can be inflated to \verb|a^b^c^b|.
Attackers can also use jump instructions to make the real execution flow different from the disassembly script, as in Fig.~\ref{fig:obfuscation}.

\begin{figure}[htbp]
    \centerline{\includegraphics[width=\linewidth]{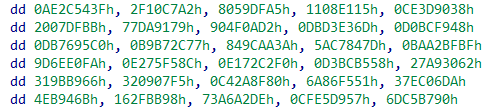}}
    \caption{IDA Pro failed to disassemble encrypted instructions}
    \label{fig:encryption}
\end{figure}

Unlike obfuscation, code encryption packs and encrypts executable files on the disk. They will decrypt themselves during execution.
It means they are nearly impossible to analyze just by static disassembly relying instead on execution and reviewing of system logs.
As shown in Fig.~\ref{fig:encryption}, IDA Pro failed to disassemble encrypted instructions and only displays hexadecimal machine code.

\subsection{The Portable Executable format}
\begin{figure}[htbp]
    \centerline{\includegraphics[width=\linewidth]{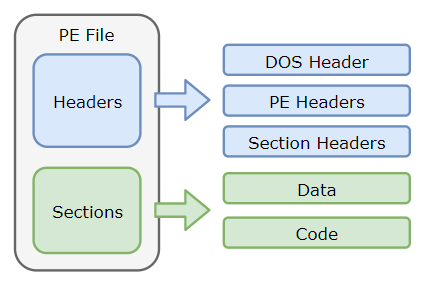}}
    \caption{Windows PE format}
    \label{fig:pe-format}
\end{figure}

The \emph{Portable Executable (PE)} format is a file format for executable files on Windows systems \cite{pe-format}, consisting of several headers and sections as in Fig.~\ref{fig:pe-format}.
Headers can be regarded as metadata, located at the beginning of an executable file.
They encapsulate system-related information, such as API export and import tables, resources (icons, images and audio, etc.), and the distribution of data and code.
This information is critical for malware analysis.
Data and executable code are stored in different sections behind headers, depending on their functions.

Analyzing the PE format does not need disassembly tools and attackers cannot completely erase these data.
So even if the malware is encrypted, most structures are still available but may not be very accurate.

\subsection{Traditional Malware Analysis}
Traditional methods include static disassembly analysis and dynamic behavior analysis.
As mentioned earlier, they heavily rely on experts.

Static analysis must disassemble machine code firstly via professional tools like IDA Pro \cite{ida-pro}.
Analysts explore control flows and instructions to understand malicious behaviors.
In theory, it provides complete code coverage, but is time-consuming and vulnerable to code obfuscation or encryption.

In contrast, dynamic analysis can effectively address code obfuscation.
Because it only focuses on system events and does not care about detailed instructions.
Moreover, it is not susceptible to encryption since encrypted files must decrypt themselves before execution.
However, dynamic analysis is costly and requires virtual environments to run the malware.
In addition, some malicious behaviors will not be recorded because the environment does not meet the execution conditions.
Malware can also detect virtual environments and hide itself.

\section{RELATED WORK}
In this section, we primarily discuss malware analysis combined with machine learning and hand-designed static features.
Since their processes can be explained in terms of these underlying, interpretable features, analysts can undertake deeper exploration according to the model output.
We also examine the small number of deep learning models which have been applied to the problem.
These utilize image and byte representations to achieve their goals.

In \cite{dynamic-fingerprint}, \citeauthor{dynamic-fingerprint} used strings, registry changes and API sequences to distinguish whether a new sample was a variant of a known sample.
They converted two samples into two vectors then calculated their similarity.
One shortcoming of this method was that accurate registry changes and API sequences must be recorded at execution, which requires costly virtual machines.
And due to the lack of enforcement conditions, not all malicious behaviors can be recorded.
In \cite{n-gram}, \citeauthor{n-gram} extracted 3-grams of byte codes and trained a binary classifier.
They selected the maximum high-information-gain grams as features, using One-hot vectors to represent samples. The similarity comparison relied on the average vectors of the malicious and benign.
When getting a new sample, they calculated its distances to two average vectors and found the nearest one.
For these two methods, a serious issue is that malware can hide signatures or fingerprints through polymorphic engines and packers with the purpose of being harder for anti-virus software to detect.
In \cite{packed-malware}, \citeauthor{packed-malware} proposed a binary classification system for packed samples.
They unpacked files and extracted PE Headers, then used the forward selection method to pick seven features.
Unlike the first two, their model did not completely rely on detailed assembly instructions and system behaviors, but introduced macroscopic information, such as the number of executable sections and the debug information flag.
However, it was unsuitable for files that could not be unpacked.

In the face of these problems which are difficult to solve by fingerprinting, \citeauthor{malware-image} proposed an innovative method in \cite{malware-image}.
They converted a malware sample's byte content into a gray-scale image. The images belonging to the same family appear similar in layout and texture.
They used $K$-Nearest Neighbors algorithm with Euclidean Distance to determine whether the samples were derived from the same origin.
Neither disassembly nor code execution was required, and simple transformations of fingerprints by polymorphic engines usually do not affect the general layout and texture of the image.
As a limitation, image representation has a problem that two malware images can be similar even if they belong to different families,
because the same visual resources are used among these samples, like icons and user interface components.
It is difficult to remove them before analysis because they are not in the same position.
And in \cite{malware-machine-learning-rise}, \citeauthor{malware-machine-learning-rise} thought image representation might produce non-existing spatial correlations between pixels in different rows.

In recent years, some deep learning models have also been applied in the field of malware classification.
In \cite{deep-cnn-malware-image}, \citeauthor{deep-cnn-malware-image} combined Convolutional Neural Networks with gray-scale image representation.
Their model architecture was based on VGG-16 \cite{deep-cnn-image-recognition} and achieved $99.97\%$ accuracy, which is the best result we have found to date.
In \cite{eat-exe}, \citeauthor{eat-exe} built a Convolutional Neural Network and used all bytes of a sample as raw data.
But instead of training the network on raw bytes, they inserted an embedding layer to map each byte to a fixed-length feature vector.
It could reduce incorrect correlations between two bytes.
In other words, the certain bytes are closer to each other than other values, which is incorrect in terms of the assembly instruction context.
And using Convolutional Neural Networks with a global max-pooling could increase the robustness when facing minor alterations in bytes.
In contrast, traditional byte $n$-gram methods are dependent on exact matches.
For deep learning models with byte-based representation, \citeauthor{malware-machine-learning-rise} thought the main advantage of such an approach is that it can be applied to samples from different systems and hardware,
because they are not affected by file formats \cite{malware-machine-learning-rise}.
However, the size of byte sequences is too large and the meaning of each byte is context-dependent. Byte-based representation does not contain this information.
Another challenge is that adjacent bytes are not always correlated because of jumps and function calls.

Apart from these models, \citeauthor{malware-machine-learning-rise} mentioned several challenges in the face of malware analysis \cite{malware-machine-learning-rise}. One of them is \emph{Concept Drift}.
In many other machine learning applications like digit classification, the mapping learned from historical data will be valid for new data in the future, and the relationship between input and output does not change over time.
But for malware, due to function updates, code obfuscation and bug fixes, the similarity between previous and future versions will degrade slowly over time, decaying the detection accuracy.
Furthermore, the interpretation of models and features should also be considered.
When an incorrect classification happens, analysts need to understand why and know how to fix it. This is challenging without clear interpretability and explainability.
Even in the absence of miss-classifications, analysts prefer to understand how a classification has been arrived at.
This is the main reason why we did not choose a deep learning model.

\section{FEATURE EXTRACTION}
In practice, malware is in the form of executable files.
But considering the limitations of dynamic analysis, we chose a static textual dataset from the 2015 Microsoft Malware Classification Challenge \cite{microsoft-malware-challenge}.
It contains 10868 malware samples representing a mix of nine families, described in Table.~\ref{tbl:malware-families}.
Each sample has two files of different forms: machine code and disassembly script generated by IDA Pro, as shown in Listing.~\ref{lst:machine-code-sample} and Listing.~\ref{lst:disassembly-sample} respectively.

\begin{table}[htbp]
    \caption{Malware families in the dataset}
    \begin{center}
        \begin{tabular}{cccc}
            \toprule
            ID & Name & \# Samples & Type \\
            \midrule
            1 & \verb|Ramnit| & 1541 & Worm \\
            2 & \verb|Lollipop| & 2478 & Adware \\
            3 & \verb|Kelihos_ver3| & 2942 & Backdoor \\
            4 & \verb|Vundo| & 475 & Trojan \\
            5 & \verb|Simda| & 42 & Backdoor \\
            6 & \verb|Tracur| & 751 & Trojan Downloader \\
            7 & \verb|Kelihos_ver1| & 398 & Backdoor\\
            8 & \verb|Obfuscator.ACY| & 1228 & Obfuscated malware \\
            9 & \verb|Gatak| & 1013 & Backdoor \\
            \bottomrule
        \end{tabular}
    \end{center}
    \label{tbl:malware-families}
\end{table}

\begin{lstlisting}[label=lst:machine-code-sample, caption=A sample's machine code snippet, basicstyle=\scriptsize\ttfamily]
10001100 C4 01 74 AC D9 EE D9 C0 DD EA DF E0 F6 C4 44 7A
10001110 0A DD D9 DD 17 DD 1E 8B E5 5D C3 DD D8 E8 74 26
10001120 00 00 DC 0D 78 65 00 10 DC 34 24 E8 60 26 00 00
10001130 DD 44 24 08 D8 C9 DD 1F DC 4C 24 10 DD 1E 8B E5
\end{lstlisting}

\begin{lstlisting}[label=lst:disassembly-sample, caption=A sample's disassembly snippet, basicstyle=\scriptsize\ttfamily]
.text:10001106 D9 C0        fld     st
.text:10001108 DD EA        fucomp  st(2)
.text:1000110A DF E0        fnstsw  ax
.text:1000110C F6 C4 44     test    ah, 44h
.text:1000110F 7A 0A        jp      short loc_1000111B
\end{lstlisting}

Without executable files, dynamic analysis cannot be conducted. All features can only be extracted from textual content.
The features described in Section.~\ref{sct:import-library}, Section.~\ref{sct:pe-section-size}, Section.~\ref{sct:pe-section-permission} and Section.~\ref{sct:content-complexity} are proposed by us.
API $n$-grams in Section.~\ref{sct:api-4-gram} is an effective feature in dynamic behavior analysis. We tested it to check if it is also applicable for static analysis.

\subsection{File Size}
\begin{figure}[htbp]
    \centerline{\includegraphics[width=\linewidth]{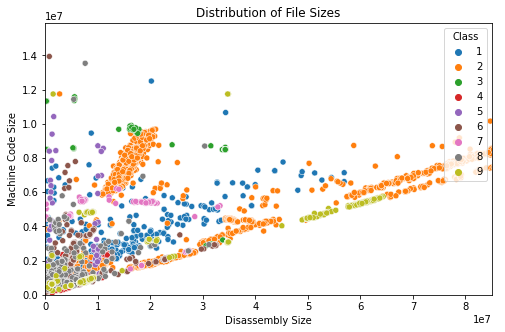}}
    \caption{The distribution of file sizes}
    \label{fig:file-sizes}
\end{figure}

The file size is the simplest feature, containing the sizes of disassembly and machine code files, and their ratios.
File sizes vary according to the functional complexity and code obfuscation of different malware families. The distribution is shown in Fig.~\ref{fig:file-sizes}.
Size ratios may represent code encryption.
If a sample is encrypted, disassembly process may fail and the ratio of its machine code size to the disassembly size will be different from the other samples.

\subsection{API 4-gram}
\label{sct:api-4-gram}
The API sequence is almost the most commonly used feature in dynamic analysis. It directly uses malicious or suspicious API sequences to classify malware.
Each malware family has distinct functions. For instance, \verb|Lollipop| is Adware showing advertisements as users browse websites.
\verb|Ramnit| can disable the system firewall and steal sensitive information. These functions rely on different APIs.

API sequences should be extracted by dynamic execution since it can reduce the negative impact of code encryption.
However, if dynamic analysis is not possible, we can only use regular expressions to match \verb|call| and \verb|jmp| instructions whose target is an import API from static disassembly scripts.
It has a huge negative impact on accuracy and we will discuss this in detail in a later section.
Finally, 402972 API 4-grams were extracted and only the 5000 most frequent items were retained.

\subsection{Opcode 4-gram}
The opcode sequence is also commonly used. It focuses on assembly instructions.
Opcodes are defined by CPU architectures, not by systems as in the case of APIs.
So they are compatible with different systems built on the same architecture.
For example, the opcode sequence of Listing.~\ref{lst:obfuscation} is \verb|jmp|, \verb|db|, \verb|mov|, \verb|add|.
The other parts of each instruction are ignored.
1408515 opcode 4-grams were extracted and only the 5000 most frequent items were retained.

\subsection{Import Library}
\label{sct:import-library}
As mentioned in Section.~\ref{sct:api-4-gram}, each malware family has distinct functions.
They must import system or third-party libraries to achieve.
So a typical machine learning feature is the number of APIs per imported library used by malware.
But API numbers would be inaccurate if malware calls an API with dynamic methods such as by \verb|GetProcAddress|.

\begin{figure}[htbp]
    \centerline{\includegraphics[width=\linewidth]{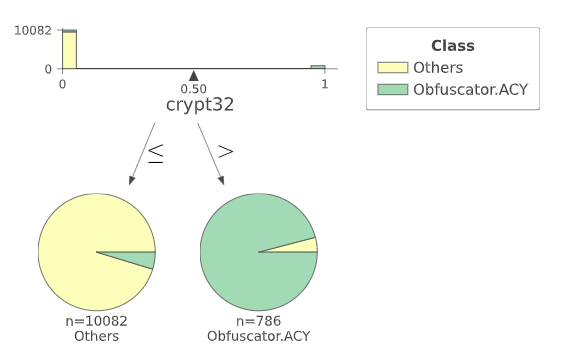}}
    \cprotect\caption{A Decision Tree for \verb|Obfuscator.ACY|}
    \label{fig:obfuscator.acy-decision-tree}
\end{figure}

We proposed a simpler variant, directly using One-Hot Encoding to indicate whether a library is imported by malware.
There are 570 different import libraries in the dataset. The 300 with the highest number of occurrences among them were retained.
Fig.~\ref{fig:obfuscator.acy-decision-tree} demonstrates how this feature distinguishes \verb|Obfuscator.ACY| from others using a cryptographic library \verb|Crypt32|.
With the list of import libraries, we can make assumptions about a sample's purpose. These assumptions may be inaccurate, but can provide references for further in-depth analysis.

\subsection{PE Section Size}
\label{sct:pe-section-size}
PE files consist of several sections. Each section stores different types of bytes.
The number of sections, their uses and attributes are defined by software development tools and programmers based on functionality.

This feature focuses on section sizes.
Each section has two types of sizes: a virtual size and a raw size.
They are \verb|VirtualSize| and \verb|SizeOfRawData| fields of structure \verb|IMAGE_SECTION_HEADER| in PE Headers, respectively.
The raw size is exactly the size of a section on the disk, and the virtual size is the size of a section when it has been loaded into memory.
For instance, a section may store only uninitialized data whose values are only available after startup.
There is no need to allocate space for it on the disk, so the raw size is zero, but the virtual size is not.
The ratio of the two types of sizes is also included in the feature.
The dataset contains 282 sections with different names.
Each section has three attributes, so the full feature has 846 dimensions.
After feature selection using Random Forest based on Gini Impurity, only the 25 most essential dimensions were retained.
Most of them are standard sections defined by software development tools as in Table.~\ref{tbl:top-sections}.

\begin{table}[htbp]
    \caption{The most important sections}
    \begin{center}
        \begin{tabular}{cc}
            \toprule
            Section & Stored Data \\
            \midrule
            \verb|text| & Executable code \\
            \verb|data| & Normal data \\
            \verb|idata| & Import libraries and functions \\
            \verb|rdata| & Read-only data \\
            \verb|bss| & Block Starting Symbols \\
            \verb|tls| & Thread Local Storage \\
            \bottomrule
        \end{tabular}
    \end{center}
    \label{tbl:top-sections}
\end{table}

\subsection{PE Section Permission}
\label{sct:pe-section-permission}
\begin{figure}[htbp]
    \centerline{\includegraphics[width=\linewidth]{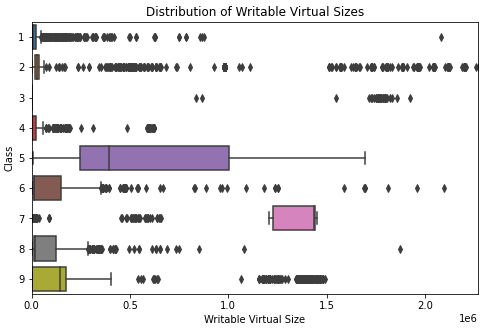}}
    \caption{The distribution of writable virtual sizes}
    \label{fig:writable-sizes}
\end{figure}

PE sections have access permissions, which are combinations of readable, writable and executable.
We calculated the total size of readable data, writable data and executable code separately for each sample.
Like the previous, each permission also has three attributes: a virtual size, a raw size and a ratio of the two sizes.
For example, Fig.~\ref{fig:writable-sizes} shows the distribution of writable virtual sizes.
Backdoor samples generally have a larger writable space, especially class 5 and class 7.
This feature can be regarded as a summary of PE section sizes and provides a more macroscopic view with only nine fixed dimensions.

Additionally, we think these two PE section features (sizes and permissions) have compatibility with Linux systems.
Linux uses the \emph{Executable and Linkable Format (ELF)}\footnote{\url{https://en.wikipedia.org/wiki/Executable_and_Linkable_Format}} for executable files.
It has similar section structures to the PE format.

\subsection{Content Complexity}
\label{sct:content-complexity}
Content complexity is a new feature type for malware classification. What we propose here has six fixed dimensions: the original sizes, compressed sizes and compression ratios of disassembly and machine code files.
We used Python's \verb|zlib| library to compress samples and recorded size changes.
This approximates function complexity, code encryption and obfuscation.
Listing.~\ref{lst:max-compression} is from the sample with the largest disassembly compression ratio of $12.8$.
It might be obfuscated with repetitive, roundabout instructions.
In contrast, Listing.~\ref{lst:min-compression} has the smallest disassembly compression ratio of $2.3$.
The disassembly process failed and IDA Pro can only output its original machine code.
This is because the sample is encrypted and packed by UPX\footnote{\url{https://upx.github.io}}, a famous open-source packer for executable files.
It is difficult to compress the encrypted content again, which causes a low compression ratio.
In addition to this, the use of complex, rare instructions can also lead to low compression ratios.

\begin{lstlisting}[label=lst:max-compression, caption=Disassembly snippet with the largest compression ratio, language={[x86masm]Assembler}]
    mov     eax, esi
    ror     eax, 8
    mov     ecx, esi
    ror     ecx, 5
    mov     edi, edx
\end{lstlisting}

\begin{lstlisting}[label=lst:min-compression, caption=Disassembly snippet with the smallest compression ratio, language={[x86masm]Assembler}]
    dd  66E5C7CBh, 0BABBBB02h, 8525F0EEh
    dd  8ED64AF8h, 0F93780DFh, 80B1924Ah
    dd  3C99EBD7h, 0AF12FC33h, 837520D0h
    dd  6953D3A9h, 0C7012601h, 0D76CAC3Ah
\end{lstlisting}

In theory, this feature can be used directly in the malware classification of any other CPU architecture and system.
It has better compatibility than others because it does not rely on any platform-related characteristics and structures.
But in practice, CPUs have different instruction sets.
For instance, Intel x86 is based on Complex Instruction Set Computing, while ARM is based on Reduced Instruction Set Computing.
It may affect classification accuracy.

\section{EXPERIMENTS}
For each feature and its combination, we used automatic machine learning library \verb|auto-sklearn| to search for the best parameters,
relying on Bayesian optimization, meta-learning and ensemble construction \cite{automated-machine-learning}.
$80\%$ of the dataset was used as a training set and \verb|auto-sklearn| evaluated models on it using 5-fold cross-validation.
The models include $K$-Nearest Neighbors, Support Vector Machine and Random Forest.
All experiments were conducted on 64-bit Ubuntu, Intel(R) Core(TM) i7-6700 CPU (3.40GHz) with 12GB RAM. Each model's parameter search process lasted up to one hour.
After \verb|auto-sklearn| had determined a model's optimal parameters, we used the remaining $20\%$ as a test set to calculate classification accuracy.
The results are shown in Table.~\ref{tbl:feature-accuracy}, sorted in increasing order of accuracy. Random Forest provided the best performance in all experiments.
For the ``Dimension'' column of some features, the numbers before and after the arrow indicate the size of the feature before and after the feature selection respectively.

\begin{table*}[htbp]
    \caption{The feature accuracy}
    \begin{center}
        \begin{tabular}{cccc}
            \toprule
            Feature(s) & Dimension & Best Accuracy \\
            \midrule
            All Features & $1812921 \rightarrow 10343$ & $0.9948$ \\
            Section Size, Section Permission, Content Complexity & $861 \rightarrow 40$ & $0.9940$ \\
            Section Size, Section Permission, Content Complexity, Import Library & $1431 \rightarrow 340$ & $0.9922$ \\
            Opcode 4-gram & $1408515 \rightarrow 5000$ & $0.9908$ \\
            File Size, API 4-gram, Opcode 4-gram & $1811490 \rightarrow 10003$ & $0.9899$ \\
            Content Complexity & $6$ & $0.9811$ \\
            Section Size & $846 \rightarrow 25$ & $0.9775$ \\
            Section Permission & $9$ & $0.9701$ \\
            Import Library & $570 \rightarrow 300$ & $0.9393$ \\
            File Size & $3$ & $0.9352$ \\
            API 4-gram & $402972 \rightarrow 5000$ & $0.5796$ \\
            \bottomrule
        \end{tabular}
    \end{center}
    \label{tbl:feature-accuracy}
\end{table*}

Among individual features, opcode 4-grams provided the highest accuracy of $99.08\%$, meaning static disassembly does not invoke many negative impacts on opcode 4-grams.
They are effective both in dynamic and static analysis, but their extraction requires much time and computational resources.
The original dimension of opcode 4-grams before feature selection is the largest (1408515).
Content complexity, PE section sizes and PE section permissions achieved $98.11\%$, $97.75\%$ and $97.01\%$ accuracy respectively, which are satisfactory considering they are low dimensional representations.
Import libraries did not perform very well, but the prediction paths generated by a Decision Tree can provide functionality comparisons between malware families, like Fig.~\ref{fig:obfuscator.acy-decision-tree}.
Other features except API sequences cannot do this.
At the beginning, we expected that the API sequence would be an effective feature in static analysis, as it does in dynamic analysis.
Unexpectedly, the API 4-gram is the worst. Its accuracy is only $57.96\%$ and involves a 5000-dimensional feature vector representation.
Our result shows that incorrect sequences extracted from static disassembly scripts have a very negative effect on feature validity.

Among integrated features, the combination of PE section sizes, PE section permissions and content complexity is almost the best with $99.40\%$ accuracy and 40 dimensions.
If all features were used, the accuracy was only improved by $0.08\%$, while the number of dimensions increased dramatically to 10343.
Additionally, the highest accuracy of $99.48\%$ we achieved is still lower than $99.97\%$ in \cite{deep-cnn-malware-image} written by \citeauthor{deep-cnn-malware-image}.
If interpretability is not considered, the combination of Convolutional Neural Networks and gray-scale images they used is obviously an excellent model.

\section{LIMITATIONS OF STATIC DISASSEMBLY}
The dataset contains only static text. In general, it negatively affects classification accuracy.
We identified three specific problems. They mainly affect the extraction of API sequences and import libraries.

\subsection{Lazy Loading}
In the process of extracting import libraries, only the libraries in the \emph{Import Table} can be extracted, which is a structure in PE Headers used to import external APIs.
These libraries will be automatically loaded when malware starts.
In order to make malicious behavior more hidden, developers can use lazy loading to load a library just before it is about to be used.
Lazily loaded libraries cannot be extracted from static content.
Table.~\ref{tbl:top-libraries} shows the top libraries based on Gini Impurity.
They are ubiquitous and have no special significance for malware classification.
A reasonable speculation is that sensitive libraries are lazily loaded and PE Headers only contain regular libraries.

\begin{table}[htbp]
    \caption{The top five important libraries}
    \begin{center}
        \begin{tabular}{cc}
            \toprule
            Library & Description \\
            \midrule
            \verb|MSASN1| & Abstract Syntax Notation One Runtime \\
            \verb|MSVCRT| & Microsoft Visual C++ Runtime \\
            \verb|UXTheme| & Microsoft Windows Controls \\
            \verb|OpenGL32| & Open Graphics Library \\
            \verb|ADVAPI32| & Microsoft Security and Registry Services \\
            \bottomrule
        \end{tabular}
    \end{center}
    \label{tbl:top-libraries}
\end{table}

\subsection{Name Mangling}
Compared with import libraries, the API sequence is more negatively affected.
We found two reasons. The first is \emph{Name Mangling}.
It allows different programming entities to be named with the same identifier, like C++ overloading.
Compilers can select the appropriate function based on parameters. It is convenient for programmers.
Internally, compilers need different names to distinguish them.
As a test, we exported two C++ overloading functions \verb|void fun(int)| and \verb|void fun(double)|.
Table.~\ref{tbl:mangled-names} shows their mangled names. They vary depending on compilers.

\begin{table}[htbp]
    \caption{Mangled overloading names}
    \begin{center}
        \begin{tabular}{ccc}
            \toprule
            Compiler & \verb|void fun(int)| & \verb|void fun(double)| \\
            \midrule
            GCC 8 & \verb|_Z3funi| & \verb|_Z3fund| \\
            Visual C++ 2022 & \verb|?fun@@YAXH@Z| & \verb|?fun@@YAXN@Z| \\
            \bottomrule
        \end{tabular}
    \end{center}
    \label{tbl:mangled-names}
\end{table}

Name mangling adds noise to the API $n$-gram extraction. For the same or similar functions, we may extract more than one name.
A theoretical solution is to convert mangled names back to the same original name.
However, in practice, it is challenging to develop converters for every possible compiler and language.
Moreover, some compilers do not disclose their detailed name mangling mechanism.

\subsection{Jump Thunk}
\emph{Jump Thunk} is the second reason for the poor performance of API sequences.
Many compilers generate a jump thunk, a small code snippet, for each external API, then convert all calls to an API into calls to its jump thunk.
This mechanism can provide an interface proxy for an API.
In Listing.~\ref{lst:jump-thunks}, there are two Windows file manipulation APIs.
After defining all jump thunks in the beginning part of the code, the rest only uses a jump thunk to call an external API.
For instance, all calls to \verb|WriteFile| became calls to its thunk \verb|j_write_file|.

\begin{lstlisting}[label=lst:jump-thunks, caption=Jump thunks, language={[x86masm]Assembler}]
    j_write_file    proc
        jmp     WriteFile
    j_write_file    endp

    j_read_file     proc
        jmp     ReadFile
    j_read_file     endp

    call    j_read_file
    call    j_write_file
    call    j_read_file
\end{lstlisting}

Jump thunks make API sequences inaccurate.
If we simply use linear scanning to extract external API calls from Listing.~\ref{lst:jump-thunks}, we will get the sequence \verb|WriteFile|, \verb|ReadFile|.
But the true sequence is \verb|ReadFile|, \verb|WriteFile|, \verb|ReadFile|, covered and hidden by their thunks.
Theoretically, we can recognize jump thunks and match them to external APIs.
But thunks' names are random and their content may be more complex than jump instructions.

\section{PRACTICAL APPLICATION}
\begin{figure}[htbp]
    \centerline{\includegraphics[width=\linewidth]{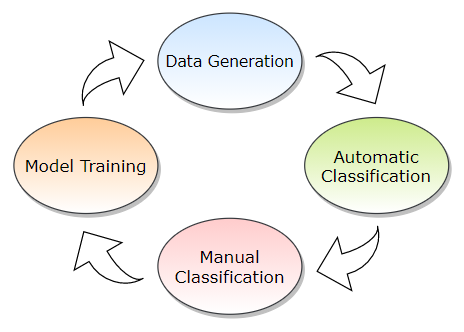}}
    \caption{The automatic malware classification workflow}
    \label{fig:workflow}
\end{figure}

As discussed in \cite{malware-machine-learning-rise}, unlike other machine learning applications like handwritten digit classification, where the shape of numbers is not updated over time,
the similarity between previous and future malware will degrade over time due to function updates and polymorphic techniques.
Polymorphic techniques can automatically and frequently change identifiable characteristics like encryption types and code distribution to make malware unrecognizable to anti-virus detection.
To solve this, we designed an automatic malware classification workflow to apply and enhance our classifier in practice with IDA Pro's Python development kit, as shown in Fig.~\ref{fig:workflow}.
The source code is available on the GitHub\footnote{\url{https://github.com/czs108/Microsoft-Malware-Classification}} and makes available as practical features the following contributions.

\begin{enumerate}
    \item \textbf{Data Generation} \\
    In general, analysts can only collect raw executable malware, not disassembly scripts like those provided in the Microsoft's dataset.
    To generate similar data, we developed an IDA Pro script that can be run from the command line with IDA Pro's parameters \verb|-A| and \verb|-S|,
    which launch IDA Pro in autonomous mode and make it run a script.
    For each executable file, it produces disassembly instructions and hexadecimal machine code, relying on IDA Pro's disassembler.
    These two output files are in the same format as the files used for training in the dataset.

    \item \textbf{Automatic Classification} \\
    We used another automatic machine learning library \verb|TPOT| to search for the best model for the feature combination of PE section sizes, PE section permissions and content complexity.
    We think this combination maintains a good balance between accuracy and the number of dimensions.
    \verb|TPOT| achieved $99.26\%$ accuracy, slightly lower than \verb|auto-sklearn| ($99.40\%$).
    Unlike \verb|auto-sklearn|, \verb|TPOT| uses Genetic Programming to optimize models \cite{tree-based-automated-machine-learning}.
    Once the search is complete, it will provide Python code for the best pipeline. \verb|auto-sklearn| does not have a similar function.
    With the fitted model, we developed an IDA Pro classifier plug-in.
    When an analyst opens an executable sample with IDA Pro, the plug-in will produce the required disassembly and machine code files, extract features and perform classification as in Listing.~\ref{lst:ida-plug-in}.

    \item \textbf{Manual Classification} \\
    Although automatic classification is very useful, the result may be inaccurate or in doubt especially when a sample does not belong to known families.
    Therefore the plugin provides the probability distribution for analysts to perform in-depth analysis manually to determine a sample's exact family.

    \item \textbf{Model Training} \\
    With sufficient output files and labels of the latest samples, the classifier can be retrained and strengthened either manually or in an automated fashion.
\end{enumerate}

\begin{lstlisting}[label=lst:ida-plug-in, caption=An output of IDA Pro classifier plug-in, language=bash]
    0.53 -> Ramnit
    0.24 -> Lollipop
    0.17 -> Obfuscator.ACY
    0.05 -> Gatak
    0.01 -> Simda
    0.01 -> Vundo
    0.00 -> Tracur
    0.00 -> Kelihos_ver1
    0.00 -> Kelihos_ver3
\end{lstlisting}

Our model was trained on these nine malware families only, so if an input sample does not belong to them, the model will get an incorrect result or classify the sample into the family that is most similar to its actual type.
In the ideal case, these features are applicable to more families if more datasets are available. We just need to retrain the mode.
But we think that as the number of malware families becomes larger, their effectiveness may gradually decrease. They may be too simple to distinguish between huge families.

\section{CONCLUSION AND FUTURE WORK}
This paper demonstrates how novel, highly discriminative features of relatively low dimensionality when combined with automatic machine learning approaches can provide highly competitive accuracy for malware classification.
Compared with traditional manual analysis, machine learning can provide a fast and accurate classifier after training on the latest malware samples.
It does not rely on an understanding of code.
Unlike API and opcode $n$-grams, which aim to match specific malicious operations, our features focus more on macroscopic information about malware.
In theory, these features are more compatible with multiple operating systems and not susceptible to code encryption.
One shortcoming is that they cannot offer a detailed understanding of malicious behaviors like API sequences.
Analysts must combine multiple features in order to perform more in-depth analysis.
In addition, the negative limitations and effects of static text are more severe than we thought, especially for API $n$-grams.
It is challenging to extract exact API sequences from disassembly scripts simply with linear scanning.

We conclude with a number of open avenues for research that might reduce the negative effects of static analysis and improve machine learning models for malware processing:

\begin{itemize}
    \item
    Remove regular libraries from the import library feature.
    Machine learning models are forced to use only sensitive libraries to classify samples. Note a potential problem here is that only a tiny number of sensitive libraries may be extracted.

    \item
    Although many C/C++ compilers exist, there are not many commonly used versions.
    We can consider developing name demangling for common compilers and renaming APIs using our defined convention.

    \item
    The core of a disassembly script is assembly instructions.
    So assemblers may be helpful to perform code analysis to determine the correspondence between APIs and jump thunks.
\end{itemize}

\bibliographystyle{IEEEtranN}
\bibliography{malware-classification}

\end{document}